\documentclass[fleqn,usenatbib]{mnras}

\usepackage{newtxtext,newtxmath}
\usepackage{xcolor}

\usepackage{orcidlink}

\usepackage[T1]{fontenc}

\DeclareRobustCommand{\VAN}[3]{#2}
\let\VANthebibliography\thebibliography
\def\thebibliography{\DeclareRobustCommand{\VAN}[3]{##3}\VANthebibliography}

\usepackage{graphicx}	
\usepackage{amsmath}	

\title{A cautionary lesson from Gaia systematics: the mono-metallic globular cluster NGC 5904}

\author[P. Bianchini \& A. Mastrobuono-Battisti]{
Paolo Bianchini,$^{1}$\thanks{E-mail: paolo.bianchini@astro.unistra.fr}\orcidlink{0000-0002-0358-4502}
Alessandra Mastrobuono-Battisti,$^{2}$ \orcidlink{0000-0002-2386-9142}
\\
$^{1}$ CNRS, Observatoire Astronomique de Strasbourg, Universit\'e de Strasbourg, UMR 7550, 11 rue de l'Universit\'e, 67000, Strasbourg, France\\
$^{2}$ GEPI, Observatoire de Paris, PSL Research University, CNRS, Place Jules Janssen, 92195, Meudon, France
}

\date{Accepted 2023 September 27. Received 2023 September 21; in original form 2023 July 28}

\pubyear{2023}

\begin{document}
\label{firstpage}
\pagerange{\pageref{firstpage}--\pageref{lastpage}}
\maketitle

\begin{abstract}
The study of the chemistry of the stellar populations in Globular Clusters (GCs) is a fundamental task to unveil their formation in the high-redshift universe and to reconstruct the build up of our Galaxy. Using metallicity estimates from BP/RP low-resolution Gaia DR3 spectra, a recent work presented the surprising detection of two stellar populations with distinct metallicities in the stellar stream of the GC NGC 5904, otherwise considered a mono-metallic system. The presence of these two populations, with [Fe/H]$\sim-1.4$ and [Fe/H]$\sim-2.0$ dex, was taken as the evidence of a merger origin of the cluster. In this Letter, using the same data set complemented by new robust metallicity estimates, we carry out a detailed analysis of the metallicity distribution of stars belonging both to the cluster and to its stellar stream, explicitly focusing on the subtle effects of data systematics. We demonstrate that the population at [Fe/H]$\sim-2.0$ dex is a data artefact due to error systematics, affecting especially faint stars. The new higher quality metallicity sample corroborates this finding, and it indicates the presence of only one population of stars with metallicity of [Fe/H]$\sim-1.3$ dex, in agreement with previous literature studies. We, therefore, conclude that both NGC 5904 and its stellar stream are mono-metallic systems, and emphasize the need of carefully examining systematic effects in large and complex data.

\end{abstract}




\section{Introduction}
Globular clusters (GCs) are massive and old stellar systems that populate the halo, disc and bulge of the Galaxy. While variations in elements involved in hot H-burning processes (i.e. light element abundances) seem to be ubiquitous in their stars \citep[see e.g. ][]{Gratton2012, Renzini2015, Bastian2018, Gratton2019}, GCs are still considered approximately mono-metallic systems with no large variations in the abundances of heavy elements \citep{Gratton2019}.

A family of massive clusters escape this definition, showing large internal metallicity variations \citep{Yong2014, Marino2015, Johnson2015, Johnson2017, Lardo2023}. Such clusters, which include $\omega$ Cen, M54, NGC 1851, Terzan 5, Liller 1 to mention a few, show broad metallicity distribution functions, possibly with different peaks, corresponding to multiple sequences in their colour-magnitude diagrams. These features indicate the presence of distinct stellar populations with different ages. The origin of such anomalies, as well as the origin of the spread in light elements, is still a matter of debate and points towards complex and extended star formation histories, at odds with what is commonly assumed for GCs.

Metallicity variations in GCs are often used to reconstruct their formation history, suggesting complex formation mechanisms. To produce such anomalies, clusters needed to be significantly more massive than they appear today, to be able to retain the metal-enriched SN type II ejecta during their early life. Therefore, a fraction of these ``metal complex'' GCs are suspected to be the remnants of nuclei of accreted satellites, as found for M54 \citep{Monaco05, Bellazzini2008, Alfaro19, Alfaro_Cuello_2020, Kacharov22} and $\omega$ Cen \citep{Freeman1993, Dinescu1999, Hughes2000, Bekki2003, Boeker2008}. Others could be the leftovers of the bricks that build the Galactic bulge, or the result of cluster-cluster mergers or mergers with clouds of enriched gas \citep[e.g. Terzan 5, Liller 1; see][]{Ferraro2009, Ferraro2016, McKenzie2018, Khoperskov2018, Mastrobuono2019, Ferraro2021, Bastian2022, Taylor2022, Crociati2023}. 

From an observational perspective, the detailed measurement of GCs metallicity distributions requires relatively high-resolution spectroscopy (e.g. R>2000), and, therefore, it remains a challenging task, often limited to the study of bright stars only. 
However, large all-sky data surveys, such as the third data release of the Gaia mission (DR3; \citealp{Gaia2016,Gaia2023}), recently opened the possibility of studying both the chemical and kinematic properties of a large amount of resolved stars in Milky Way GCs. These large data sets often use different methodologies and require in-depth comparisons and the study of possible data systematics (e.g. \citealp{Soubiran2022, Martin2023}). In particular, Gaia DR3 provides stellar parameters, including metallicities [M/H], of about 470 million stars, obtained from low resolution BP/RP spectra using the General Stellar Parameterizer from Photometry (GSP-Phot, \citealp{Andrae2023a}).These measurements have the advantage of sampling in a homogeneous way both bright and faint stars (down to magnitudes G$\sim$20), however, as the authors point out, the GSP-Phot metallicity estimates are dominated by large systematic errors. Without a suitable calibration, \cite{Andrae2023a} advise against their use for quantitative analysis. As a solution to this limitation, they provide an empirical calibration of their [M/H] estimates to the [Fe/H] scale of LAMOST DR6.

Using these calibrated  GSP-Phot metallicity estimates, \citet{Piatti2023} recently reported the detection of two distinct populations in the stellar stream associated to the GC NGC 5904: one population with [Fe/H]$\sim-1.4$ dex and one population with [Fe/H]$\sim-2.0$ dex. This is in strong contrast with the properties of the stars member of the GC itself, which are known to have a single metallicity value of [Fe/H]$\sim-1.34$ dex \citep{Carretta2009}, with a small spread of 0.04 dex \citep{Bailin_2019}. 

The detection of these two distinct populations in NGC 5904 stellar stream was taken as the evidence of an accreted origin of the GC and of its merger with a distinct GC responsible for the metal-poor population of [Fe/H]$\sim-2.0$ dex \citep{Piatti2023}. This conclusion was reached from the analysis of 25 stars in the stream, and it was not further validated by the detailed study of the metallicities of the stars in the GC itself. Moreover, the stars used in the analysis are all faint main sequence stars (G>18), and therefore they are in the challenging low-luminosity regime where subtle systematics could even more significantly affect the GSP-Phot metallicity estimates.

In this Letter, we aim at exploring in a coherent way both the metallicity distribution of NGC 5904 stars and of its stellar stream, while robustly identifying possible data systematics. For this purpose, we exploit both the GSP-Phot metallicity measurements used in \citet{Piatti2023} and the new metallicity estimates recently reported in \cite{Andrae2023b}, obtained from the BP/RP Gaia spectra using a robust data-driven approach, namely the XGBoost algorithm.
Comparing the two data sets, we demonstrate that NGC 5904 is composed of a mono-metallic stellar population of [Fe/H]$\sim-1.3$ dex, consistent with previous literature measurements, and that the GSP-Phot metallicity estimates are affected by strong systematics that can mimic a second population at  [Fe/H]$\sim-2.0$ dex. This strongly indicates that the detection of the metal-poor population in the tidal tails by \citet{Piatti2023} is an artefact of the GSP-Phot metallicity sample, possibly connected to the low luminosity of the stars employed in the analysis.


\section{Data analysis}
We split the analysis is two parts: in Section \ref{subsec:1}, we analyse the metallicity distribution of the member stars of NGC 5904, while in Section \ref{subsec:2}, we concentrate on the stars which are part of the tidal tails associated to the cluster. In both cases, we conduct the analysis using simultaneously the two metallicity samples derived from Gaia DR3 BP/RP spectra:
\begin{enumerate}
\item the GSP-Phot metallicity sample from Gaia DR3 (non-calibrated and calibrated according to the prescription of \citealp{Andrae2023a})\footnote{\url{https://www.cosmos.esa.int/web/gaia/dr3-gspphot-metallicity-calibration}},
\item the metallicity sample derived with the XGBoost algorithm, trained on stellar parameters from APOGEE (\citealp{Andrae2023b}; catalogue available in \citealp{Andrae2023_dataset}).
\end{enumerate}

\subsection{The metallicity of the cluster NGC 5904}
\label{subsec:1}
\begin{figure*}
	\includegraphics[width=0.8\textwidth]{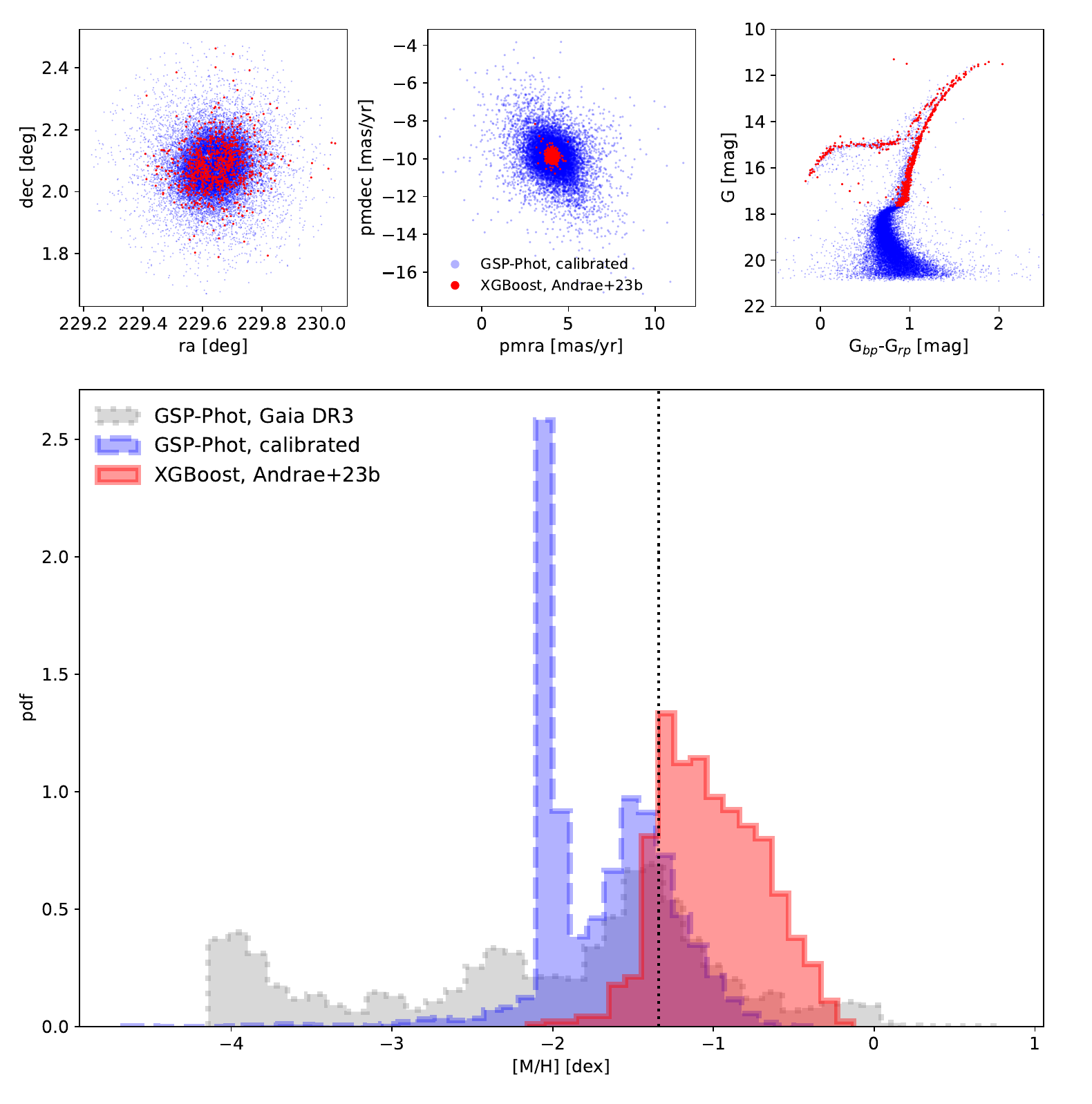}
    \caption{\textit{Top panels:} Gaia DR3 member stars of the GC NGC 5904, selected around 30 arcmin of the cluster's centre, following the membership criteria of \citet{Vasiliev2021}. From left to right, we show their spatial, velocity, and colour-magnitude distributions. Blue and red points refer to stars with GSP-Phot and XGBoost metallicity measurements, respectively. \textit{Bottom panel:} Metallicity distribution of the stars in NGC 5904. In grey we show the non-calibrated GSP-phot metallicity estimates, in blue the data calibrated as in \citet{Andrae2023a}, and in red the XGBoost metallicites. The dotted line indicates the literature value, [Fe/H]$=-1.34$ dex \citep{Carretta2009}. The double peak visible in the GSP-Phot data is not confirmed by the higher quality XGBoost sample. NGC 5904 appears to be mono-metallic, without the presence of any second population with distinct metallicity.}
    \label{fig:1}
\end{figure*}

We select cluster members using the membership provided in \citet{Vasiliev2021} based on Gaia Early Data Release 3 kinematic and photometric data. We restrict the analysis to stars within 30 arcmin from the cluster's centre and with probability memberships higher than 0.9, obtaining a total of 21\,287 stars. For all of these stars, we have GSP-Phot metallicity estimates (and the corresponding calibrated measurements) and for a subsample of 886 stars we also have XGBoost metallicities. The stars in the GSP-Phot and XGBoost samples overlap both spatially in the field of view of the cluster and in the colour-magnitude diagram. The XGBoost sample is composed of brighter stars, with magnitudes $G<18$. These two data samples are displayed in the top panels of Figure \ref{fig:1} as blue and red points, respectively, in physical, velocity, and colour-magnitude spaces.

In the bottom panel of Figure \ref{fig:1} we plot the metallicity distribution of our samples: in grey we indicate the non-calibrated GSP-Phot metallicities, in blue and in red the calibrated GSP-Phot and the XGBoost metallicities, respectively. From this figure, it is apparent that the calibration of GSP-Phot metallicities is essential to recover metallicity values consistent with previous literature measurements of NGC 5904, shown as the dotted vertical line at [Fe/H]$\sim-1.34$ dex \citep{Carretta2009}. The calibrated measurements display a non-homogeneous distribution, with a double peak: one at $\sim-1.4$ dex and one at $\sim-2.0$ dex. These two values correspond to the two populations reported in \citet{Piatti2023} for the tidal tails of the cluster (we will further address this point in Section \ref{subsec:2}). On the contrary, the metallicity distribution from the more robust XGBoost measurements does not show a double peak, but rather a mono-metallic distribution with a peak around $\sim-1.3$ dex, fully consistent with the literature values. This discrepancy can be taken as an indication of the presence of strong systematic errors in the GSP-Phot metallicities estimates, as already cautioned by \citet{Andrae2023a}.


\begin{figure*}
	\includegraphics[width=0.8\textwidth]{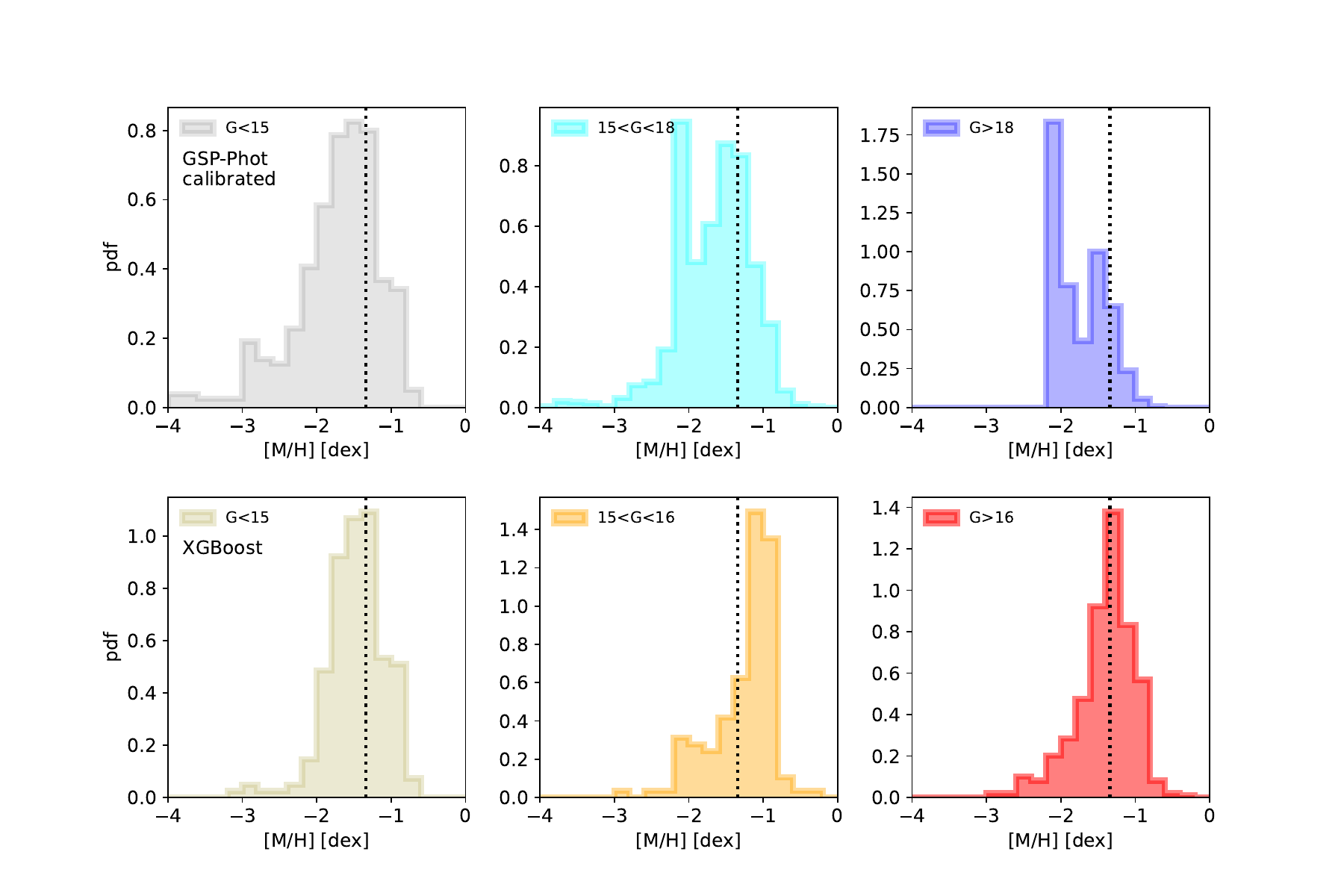}
    \caption{Metallicity distribution for different magnitude bins for stars in the GC NGC 5904, for the calibrated GSP-Phot sample (\textit{top panels}) and for the XGBoost sample (\textit{bottom panels}). The literature value of the metallicity of the cluster is shown as a dotted vertical line. For bright stars (G<15), the GSP-Phot and XGBoost samples display a similar distribution, consistent with the literature. For fainter stars, GSP-Phot metallicities display a second prominent peak at [Fe/H]$=-2.0$ dex, which is an artefact due to data error systematics, not present in the more robust XGBoost data.}
    \label{fig:2}
\end{figure*}

In order to further assess the robustness of the two samples against systematics, we plot in Figure \ref{fig:2} the metallicity distributions for stars in different magnitudes bins. For GSP-Phot metallicities, we select magnitudes bins of G<15, 15<G<18, and G>18, while for the XGBoost measurements, we select magnitudes bins of G<15, 15<G<16, and G>16. It is evident from the plots in the first row that the peak at $-2.0$ dex of the GSP-Phot sample becomes more and more apparent for faint stars (fainter than G$\sim$15), and it is a prominent feature for stars with G magnitudes fainter than 18. On the contrary, for stars brighter than G$\sim$15, the metallicity distribution is more homogeneous and is fully compatible with the one from the XGBoost sample. Moreover, for XGBoost data, the metallicity distribution does not show significant dependencies with stellar magnitudes. We take this as an argument for the robustness of the sample and conclude that the feature observed at $-2.0$ dex is very likely due to magnitude-dependent systematic errors dominating the GSP-Phot metallicity estimates.

\begin{figure*}
	\includegraphics[width=0.9\textwidth]{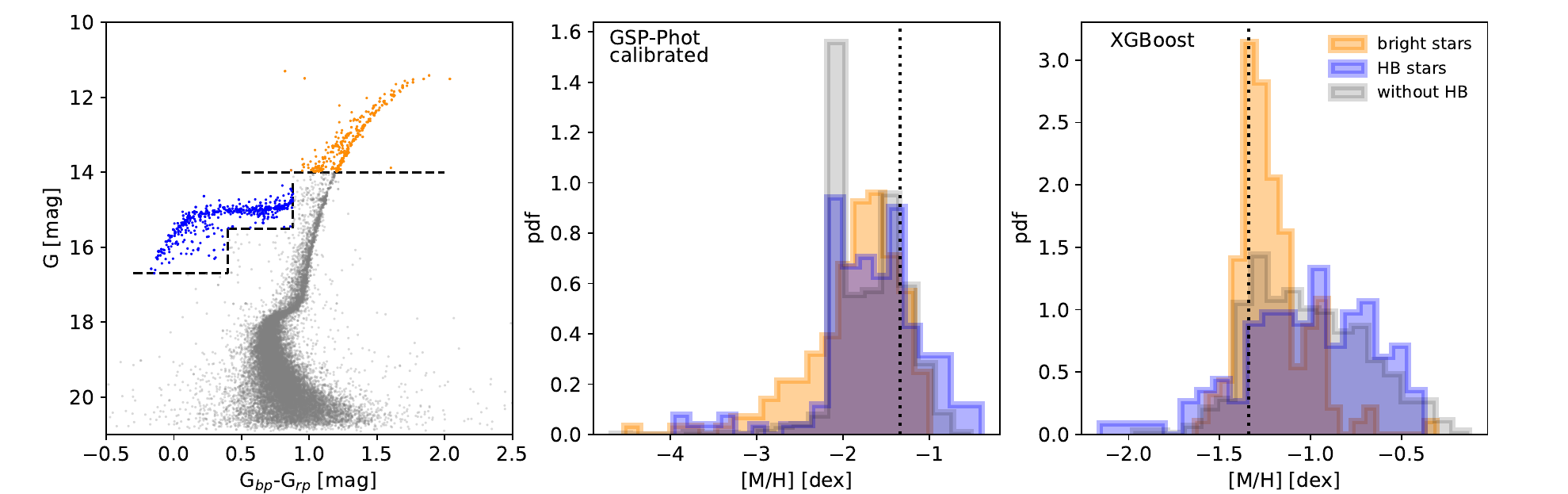}
    \caption{Metallicity distribution of bright red giant branch stars (G<14, orange symbols), horizontal branch stars (HB, blue symbols), and the entire sample without HB stars (grey symbols), for the calibrated GSP-Phot sample (\textit{central panel}) and  the XGBoost sample (\textit{right panel}). \textit{Left panel:} colour-magnitude diagram with the selection criteria for the sub-samples. HB stars show a similar metallicity distribution to that of the remaining stars, with a marginally larger scatter. Bright red giant stars display a narrower distribution consistent with literature values, for both the GSP-Phot and the XGBoost samples.}
    \label{fig:3}
\end{figure*}

Finally, we test the dependence of the metallicity estimates on the stellar type, in particular for horizontal branch (HB) stars which are known to be complex for spectroscopic modeling. In Figure \ref{fig:3}, we compare the distributions of HB stars, bright red giant branch stars (G<14) and the entire sample excluding HB stars. We do not observe strong discrepancies between HB stars and the rest of the stars, beside a larger scatter for the HB metallicity distribution. On the contrary the bright red giant stars show a narrower distribution in agreement with literature values. We conclude, consistently with the results from Figure \ref{fig:2}, that stellar luminosity has a stronger impact on the data quality rather than stellar type.

\subsection{The metallicity of the tidal tails of NGC 5904}
\label{subsec:2}

\begin{figure}
	\includegraphics[width=0.9\columnwidth]{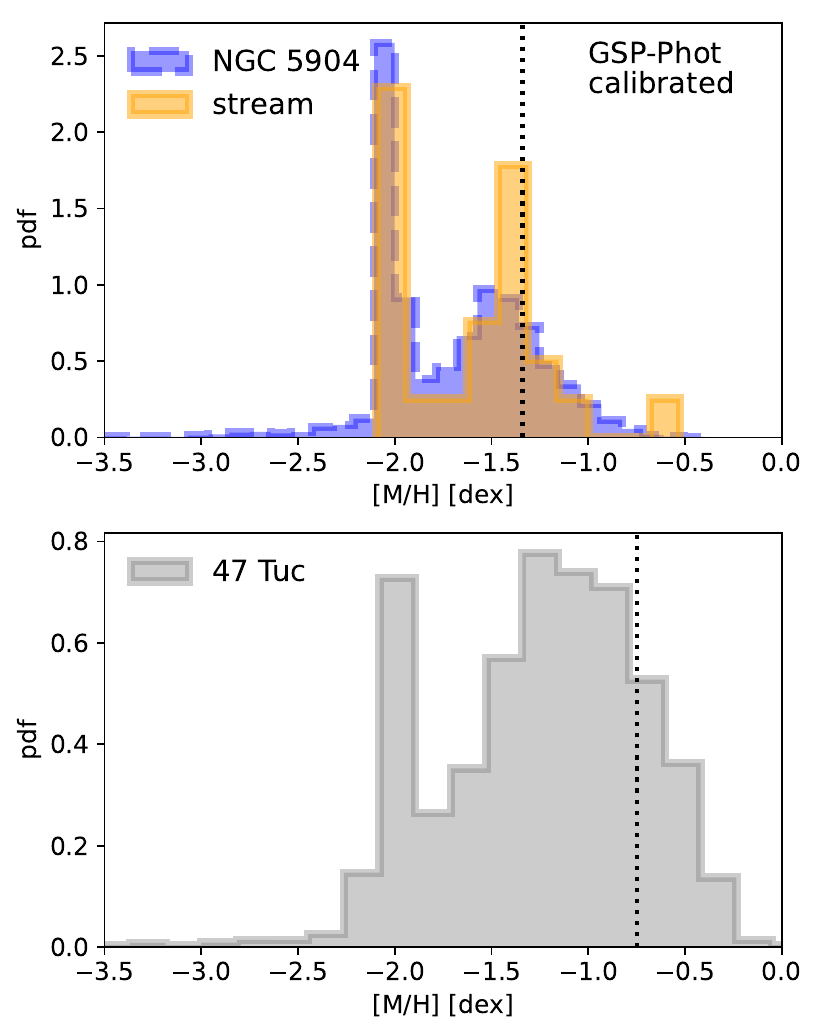}
    \caption{\textit{Top panel:} Comparison between the calibrated GSP-Phot metallicity distribution of the member stars of NGC 5904 and the stars associated to its stellar stream. The striking correspondence between the two distributions points to the presence of error systematics also in the faint stars of the tidal tails, creating a spurious peak at $-2.0$ dex. \textit{Bottom panel:} Calibrated GSP-Phot metallicty distribution for the metal richer GC 47 Tuc ([Fe/H]=$-0.75$ dex, dashed line) also showing the spurious peak at $-2.0$ dex.}
    \label{fig:4}
\end{figure}

We select stars belonging to the stellar stream of NGC 5904 following the procedure described in \citet{Piatti2023}, based on the 50 highest-ranked stream member candidates reported in \citet{Grillmair2019} and fully consistent with the stream detected in \citet{Ibata2021}. All of these stars are faint stars with $G>18$. Only a total of 25 stars have GSP-Phot metallicity estimates, while none of these stars have XGBoost data available, due to the magnitude limit of the sample. In Figure \ref{fig:3}, we compare the metallicity distribution of the stars in the stream with the one for the cluster, using only GSP-Phot calibrated values. It is evident that the stars in the stream follow the same metallicity distribution as the stars in the cluster, with a prominent feature at [Fe/H]$\sim-2.0$ dex. As demonstrated above, this feature, not observed in the higher-quality XGBoost data, is likely due to data systematics affecting GSP-Phot metallicities in the low luminosity regime. For this reason, in accordance with our analysis in Section \ref{subsec:2}, we strongly argue that the population at [Fe/H]$\sim-2.0$ dex is a data artefact also for the stellar stream stars.

To support this argument, we carry out the above analysis on a different cluster, 47 Tuc. This cluster is closer-by than NGC 5904, with a distance of d$_\odot$=4.5 kpc (7.5 kpc for NGC 5904; \citealp{Harris1996}, 2010 edition), it is relatively metal-rich, [Fe/H]=-0.75 dex \citep{Carretta2009}, and it has a small metal spread of 0.03 dex \citep{Bailin_2019}. We select a sample of 68\,244 stars using the membership criteria of \citet{Vasiliev2021}.
Even for this cluster, the distribution of GSP-Phot calibrated metallicities shows a double peak with a prominent feature at -2.0 dex (see figure \ref{fig:4}). We take this as a conclusive evidence of the general presence of data systematics creating this low metallicity feature.

\section{Discussion}
In this Letter, we carried out a homogeneous analysis of the metallicity distribution of both the cluster NGC 5904 and its associated stellar stream \citep{Grillmair2019,Ibata2021}. Using metallicity estimated derived from Gaia BP/RP spectroscopy, namely GSP-Phot metallicities and the higher quality XGBoost metallicities, we showed that: 
\begin{itemize}
    \item For stars belonging to NGC 5904, GSP-Phot data display a double peak metallicity distribution ([Fe/H]$\sim-1.4$ dex and $\sim-2.0$ dex). The shape of the distribution is not consistent with the higher quality XGBoost data, which display a mono-metallic behaviour, consistent with previous works in the literature. The feature at [Fe/H]$\sim-2.0$ dex in the data becomes more prominent for faint stars (G>15), and, therefore, it appears to be an artefact due to data systematics. 
    \item The stream of the cluster is characterized by a similar double-peaked metallicity distribution in the GSP-Phot data; however, given the low luminosity of the stars, we cannot carry out a direct comparison with the XGBoost sample. Given the magnitude dependency of the metallicity distribution for the stars in the cluster (Figure \ref{fig:2} and \ref{fig:3}), and the mono-metallicity behaviour of XGBoost data for the cluster itself, it is very unlikely that this [Fe/H]$\sim-2.0$ dex feature is a real property of the stellar population of the stellar stream.
    \item As a sanity check, we showed that another cluster, 47 Tuc, selected to be metal richer than NGC 5904, also shows a prominent peak at [Fe/H]$\sim-2.0$ dex in GSP-Phot data. We conclude that this feature is a general systematic effect present in the data set.
\end{itemize}

All the above demonstrates that GSP-Phot metallicity estimates are dominated by systematics and therefore should not be used to assess the metallicity distribution of GC stars. Our work indicates that NGC 5904 and its stellar stream are characterized by a mono-metallic stellar population, in strong disagreement with the claimed detection of two distinct stellar populations by \citet{Piatti2023}. Our result also confutes the proposed scenario according to which NGC 5904 underwent a merger with another metal-poor GC. 

We note that the quality of the data used in this work, derived from low-resolution Gaia BP/RP spectroscopy, is not of high enough quality to assess the spread in the metallicity distribution of the cluster (XGBoost metallicity estimates have typical errors of 0.2-0.3 dex; \citealp{Andrae2023b}). Therefore, we do not make any assessments of the detailed shape of the metallicity distribution of NGC 5904. Finally, our analysis shows that the metallicty estimates for both GSP-Phot and XGBoost data do not show major dependency on the stellar type; in particular we demonstrate that HB stars follow the global metallicity distribution, while bright red giant stars display a narrow distribution consistent with previous works, indicative of the higher quality of this sub-sample.

With the advent of upcoming large spectroscopic surveys, such as WEAVE and 4MOST, and the quality improvements of future Gaia data releases, we expect valuable new insights on the nature of the metallicity distribution of GCs and their tidal tails.

\section*{Acknowledgements}
PB and AMB thank the referee for the useful comments. PB acknowledges finantial support form the European Research Council (ERC) under the European Unions Horizon 2020 research and innovation programme (grant agreement No. 834148).
AMB acknowledges funding from the European Union’s Horizon 2020 research and innovation programme under the Marie Sk\l{}odowska-Curie (grant agreement No. 895174).

This work has made use of data from the European Space Agency (ESA) mission
{\it Gaia} (\url{https://www.cosmos.esa.int/gaia}), processed by the {\it Gaia}
Data Processing and Analysis Consortium (DPAC,
\url{https://www.cosmos.esa.int/web/gaia/dpac/consortium}). Funding for the DPAC
has been provided by national institutions, in particular the institutions
participating in the {\it Gaia} Multilateral Agreement.

\section*{Data Availability}
All the data used in this work are part of publicly available catalogues, as specified in the text.



\bibliographystyle{mnras}
\bibliography{example} 

\begin{thebibliography}{}
\makeatletter
\relax
\def\mn@urlcharsother{\let\do\@makeother \do\$\do\&\do\#\do\^\do\_\do\%\do\~}
\def\mn@doi{\begingroup\mn@urlcharsother \@ifnextchar [ {\mn@doi@}
  {\mn@doi@[]}}
\def\mn@doi@[#1]#2{\def\@tempa{#1}\ifx\@tempa\@empty \href
  {http://dx.doi.org/#2} {doi:#2}\else \href {http://dx.doi.org/#2} {#1}\fi
  \endgroup}
\def\mn@eprint#1#2{\mn@eprint@#1:#2::\@nil}
\def\mn@eprint@arXiv#1{\href {http://arxiv.org/abs/#1} {{\tt arXiv:#1}}}
\def\mn@eprint@dblp#1{\href {http://dblp.uni-trier.de/rec/bibtex/#1.xml}
  {dblp:#1}}
\def\mn@eprint@#1:#2:#3:#4\@nil{\def\@tempa {#1}\def\@tempb {#2}\def\@tempc
  {#3}\ifx \@tempc \@empty \let \@tempc \@tempb \let \@tempb \@tempa \fi \ifx
  \@tempb \@empty \def\@tempb {arXiv}\fi \@ifundefined
  {mn@eprint@\@tempb}{\@tempb:\@tempc}{\expandafter \expandafter \csname
  mn@eprint@\@tempb\endcsname \expandafter{\@tempc}}}

\bibitem[\protect\citeauthoryear{{Alfaro-Cuello} et~al.,}{{Alfaro-Cuello}
  et~al.}{2019}]{Alfaro19}
{Alfaro-Cuello} M.,  et~al., 2019, \mn@doi [\apj] {10.3847/1538-4357/ab1b2c},
  \href {https://ui.adsabs.harvard.edu/abs/2019ApJ...886...57A} {886, 57}

\bibitem[\protect\citeauthoryear{{Alfaro-Cuello} et~al.,}{{Alfaro-Cuello}
  et~al.}{2020}]{Alfaro_Cuello_2020}
{Alfaro-Cuello} M.,  et~al., 2020, \mn@doi [\apj] {10.3847/1538-4357/ab77bb},
  \href {https://ui.adsabs.harvard.edu/abs/2020ApJ...892...20A} {892, 20}

\bibitem[\protect\citeauthoryear{Andrae, Rix  \& Chandra}{Andrae
  et~al.}{2023a}]{Andrae2023_dataset}
Andrae R.,  Rix H.-W.,   Chandra V.,  2023a, {Robust Data-driven Metallicities
  for 120 Million Stars from Gaia XP Spectra}, \mn@doi{10.5281/zenodo.7599789},
  \url {https://doi.org/10.5281/zenodo.7599789}

\bibitem[\protect\citeauthoryear{{Andrae}, {Rix}  \& {Chandra}}{{Andrae}
  et~al.}{2023b}]{Andrae2023b}
{Andrae} R.,  {Rix} H.-W.,   {Chandra} V.,  2023b, \mn@doi [\apjs]
  {10.3847/1538-4365/acd53e}, \href
  {https://ui.adsabs.harvard.edu/abs/2023ApJS..267....8A} {267, 8}

\bibitem[\protect\citeauthoryear{{Andrae} et~al.,}{{Andrae}
  et~al.}{2023c}]{Andrae2023a}
{Andrae} R.,  et~al., 2023c, \mn@doi [\aap] {10.1051/0004-6361/202243462},
  \href {https://ui.adsabs.harvard.edu/abs/2023A&A...674A..27A} {674, A27}

\bibitem[\protect\citeauthoryear{Bailin}{Bailin}{2019}]{Bailin_2019}
Bailin J.,  2019, \mn@doi [The Astrophysical Journal Supplement Series]
  {10.3847/1538-4365/ab4812}, 245, 5

\bibitem[\protect\citeauthoryear{{Bastian} \& {Lardo}}{{Bastian} \&
  {Lardo}}{2018}]{Bastian2018}
{Bastian} N.,  {Lardo} C.,  2018, \mn@doi [\araa]
  {10.1146/annurev-astro-081817-051839}, \href
  {https://ui.adsabs.harvard.edu/abs/2018ARA&A..56...83B} {56, 83}

\bibitem[\protect\citeauthoryear{{Bastian} \& {Pfeffer}}{{Bastian} \&
  {Pfeffer}}{2022}]{Bastian2022}
{Bastian} N.,  {Pfeffer} J.,  2022, \mn@doi [\mnras] {10.1093/mnras/stab3081},
  \href {https://ui.adsabs.harvard.edu/abs/2022MNRAS.509..614B} {509, 614}

\bibitem[\protect\citeauthoryear{{Bekki} \& {Freeman}}{{Bekki} \&
  {Freeman}}{2003}]{Bekki2003}
{Bekki} K.,  {Freeman} K.~C.,  2003, \mn@doi [\mnras]
  {10.1046/j.1365-2966.2003.07275.x}, \href
  {https://ui.adsabs.harvard.edu/abs/2003MNRAS.346L..11B} {346, L11}

\bibitem[\protect\citeauthoryear{{Bellazzini} et~al.,}{{Bellazzini}
  et~al.}{2008}]{Bellazzini2008}
{Bellazzini} M.,  et~al., 2008, \mn@doi [\aj] {10.1088/0004-6256/136/3/1147},
  \href {http://adsabs.harvard.edu/abs/2008AJ....136.1147B} {136, 1147}

\bibitem[\protect\citeauthoryear{{B{\"o}ker}}{{B{\"o}ker}}{2008}]{Boeker2008}
{B{\"o}ker} T.,  2008, \mn@doi [\apjl] {10.1086/527033}, \href
  {https://ui.adsabs.harvard.edu/abs/2008ApJ...672L.111B} {672, L111}

\bibitem[\protect\citeauthoryear{{Carretta}, {Bragaglia}, {Gratton}, {D'Orazi}
  \& {Lucatello}}{{Carretta} et~al.}{2009}]{Carretta2009}
{Carretta} E.,  {Bragaglia} A.,  {Gratton} R.,  {D'Orazi} V.,   {Lucatello} S.,
   2009, \mn@doi [\aap] {10.1051/0004-6361/200913003}, \href
  {https://ui.adsabs.harvard.edu/abs/2009A&A...508..695C} {508, 695}

\bibitem[\protect\citeauthoryear{{Crociati} et~al.,}{{Crociati}
  et~al.}{2023}]{Crociati2023}
{Crociati} C.,  et~al., 2023, \mn@doi [\apj] {10.3847/1538-4357/acd382}, \href
  {https://ui.adsabs.harvard.edu/abs/2023ApJ...951...17C} {951, 17}

\bibitem[\protect\citeauthoryear{{Dinescu}, {Girard}  \& {van
  Altena}}{{Dinescu} et~al.}{1999}]{Dinescu1999}
{Dinescu} D.~I.,  {Girard} T.~M.,   {van Altena} W.~F.,  1999, \mn@doi [\aj]
  {10.1086/300807}, \href
  {https://ui.adsabs.harvard.edu/abs/1999AJ....117.1792D} {117, 1792}

\bibitem[\protect\citeauthoryear{{Ferraro} et~al.,}{{Ferraro}
  et~al.}{2009}]{Ferraro2009}
{Ferraro} F.~R.,  et~al., 2009, \mn@doi [\nat] {10.1038/nature08581}, \href
  {https://ui.adsabs.harvard.edu/abs/2009Natur.462..483F} {462, 483}

\bibitem[\protect\citeauthoryear{{Ferraro}, {Massari}, {Dalessandro},
  {Lanzoni}, {Origlia}, {Rich}  \& {Mucciarelli}}{{Ferraro}
  et~al.}{2016}]{Ferraro2016}
{Ferraro} F.~R.,  {Massari} D.,  {Dalessandro} E.,  {Lanzoni} B.,  {Origlia}
  L.,  {Rich} R.~M.,   {Mucciarelli} A.,  2016, \mn@doi [\apj]
  {10.3847/0004-637X/828/2/75}, \href
  {https://ui.adsabs.harvard.edu/abs/2016ApJ...828...75F} {828, 75}

\bibitem[\protect\citeauthoryear{{Ferraro} et~al.,}{{Ferraro}
  et~al.}{2021}]{Ferraro2021}
{Ferraro} F.~R.,  et~al., 2021, \mn@doi [Nature Astronomy]
  {10.1038/s41550-020-01267-y}, \href
  {https://ui.adsabs.harvard.edu/abs/2021NatAs...5..311F} {5, 311}

\bibitem[\protect\citeauthoryear{{Freeman}}{{Freeman}}{1993}]{Freeman1993}
{Freeman} K.~C.,  1993, in {Smith} G.~H.,  {Brodie} J.~P.,  eds,  Astronomical
  Society of the Pacific Conference Series Vol. 48, The Globular Cluster-Galaxy
  Connection. p.~608

\bibitem[\protect\citeauthoryear{{Gaia Collaboration} et~al.,}{{Gaia
  Collaboration} et~al.}{2016}]{Gaia2016}
{Gaia Collaboration} et~al., 2016, \mn@doi [\aap]
  {10.1051/0004-6361/201629272}, \href
  {https://ui.adsabs.harvard.edu/abs/2016A&A...595A...1G} {595, A1}

\bibitem[\protect\citeauthoryear{{Gaia Collaboration} et~al.,}{{Gaia
  Collaboration} et~al.}{2023}]{Gaia2023}
{Gaia Collaboration} et~al., 2023, \mn@doi [\aap]
  {10.1051/0004-6361/202243940}, \href
  {https://ui.adsabs.harvard.edu/abs/2023A&A...674A...1G} {674, A1}

\bibitem[\protect\citeauthoryear{{Gratton}, {Carretta}  \&
  {Bragaglia}}{{Gratton} et~al.}{2012}]{Gratton2012}
{Gratton} R.~G.,  {Carretta} E.,   {Bragaglia} A.,  2012, \mn@doi [\aapr]
  {10.1007/s00159-012-0050-3}, \href
  {https://ui.adsabs.harvard.edu/abs/2012A&ARv..20...50G} {20, 50}

\bibitem[\protect\citeauthoryear{{Gratton}, {Bragaglia}, {Carretta}, {D'Orazi},
  {Lucatello}  \& {Sollima}}{{Gratton} et~al.}{2019}]{Gratton2019}
{Gratton} R.,  {Bragaglia} A.,  {Carretta} E.,  {D'Orazi} V.,  {Lucatello} S.,
   {Sollima} A.,  2019, \mn@doi [\aapr] {10.1007/s00159-019-0119-3}, \href
  {https://ui.adsabs.harvard.edu/abs/2019A&ARv..27....8G} {27, 8}

\bibitem[\protect\citeauthoryear{{Grillmair}}{{Grillmair}}{2019}]{Grillmair2019}
{Grillmair} C.~J.,  2019, \mn@doi [\apj] {10.3847/1538-4357/ab441d}, \href
  {https://ui.adsabs.harvard.edu/abs/2019ApJ...884..174G} {884, 174}

\bibitem[\protect\citeauthoryear{{Harris}}{{Harris}}{1996}]{Harris1996}
{Harris} W.~E.,  1996, \mn@doi [\aj] {10.1086/118116}, \href
  {https://ui.adsabs.harvard.edu/abs/1996AJ....112.1487H} {112, 1487}

\bibitem[\protect\citeauthoryear{{Hughes} \& {Wallerstein}}{{Hughes} \&
  {Wallerstein}}{2000}]{Hughes2000}
{Hughes} J.,  {Wallerstein} G.,  2000, \mn@doi [\aj] {10.1086/301241}, \href
  {https://ui.adsabs.harvard.edu/abs/2000AJ....119.1225H} {119, 1225}

\bibitem[\protect\citeauthoryear{{Ibata} et~al.,}{{Ibata}
  et~al.}{2021}]{Ibata2021}
{Ibata} R.,  et~al., 2021, \mn@doi [\apj] {10.3847/1538-4357/abfcc2}, \href
  {https://ui.adsabs.harvard.edu/abs/2021ApJ...914..123I} {914, 123}

\bibitem[\protect\citeauthoryear{{Johnson}, {Rich}, {Pilachowski}, {Caldwell},
  {Mateo}, {Bailey}  \& {Crane}}{{Johnson} et~al.}{2015}]{Johnson2015}
{Johnson} C.~I.,  {Rich} R.~M.,  {Pilachowski} C.~A.,  {Caldwell} N.,  {Mateo}
  M.,  {Bailey} John~I. I.,   {Crane} J.~D.,  2015, \mn@doi [\aj]
  {10.1088/0004-6256/150/2/63}, \href
  {https://ui.adsabs.harvard.edu/abs/2015AJ....150...63J} {150, 63}

\bibitem[\protect\citeauthoryear{{Johnson}, {Caldwell}, {Rich}  \&
  {Walker}}{{Johnson} et~al.}{2017}]{Johnson2017}
{Johnson} C.~I.,  {Caldwell} N.,  {Rich} R.~M.,   {Walker} M.~G.,  2017,
  \mn@doi [\aj] {10.3847/1538-3881/aa86ac}, \href
  {https://ui.adsabs.harvard.edu/abs/2017AJ....154..155J} {154, 155}

\bibitem[\protect\citeauthoryear{{Kacharov} et~al.,}{{Kacharov}
  et~al.}{2022}]{Kacharov22}
{Kacharov} N.,  et~al., 2022, \mn@doi [\apj] {10.3847/1538-4357/ac9280}, \href
  {https://ui.adsabs.harvard.edu/abs/2022ApJ...939..118K} {939, 118}

\bibitem[\protect\citeauthoryear{{Khoperskov}, {Mastrobuono-Battisti}, {Di
  Matteo}  \& {Haywood}}{{Khoperskov} et~al.}{2018}]{Khoperskov2018}
{Khoperskov} S.,  {Mastrobuono-Battisti} A.,  {Di Matteo} P.,   {Haywood} M.,
  2018, \mn@doi [\aap] {10.1051/0004-6361/201833534}, \href
  {https://ui.adsabs.harvard.edu/abs/2018A&A...620A.154K} {620, A154}

\bibitem[\protect\citeauthoryear{{Lardo}, {Salaris}, {Cassisi}, {Bastian},
  {Mucciarelli}, {Cabrera-Ziri}  \& {Dalessandro}}{{Lardo}
  et~al.}{2023}]{Lardo2023}
{Lardo} C.,  {Salaris} M.,  {Cassisi} S.,  {Bastian} N.,  {Mucciarelli} A.,
  {Cabrera-Ziri} I.,   {Dalessandro} E.,  2023, \mn@doi [\aap]
  {10.1051/0004-6361/202245090}, \href
  {https://ui.adsabs.harvard.edu/abs/2023A&A...669A..19L} {669, A19}

\bibitem[\protect\citeauthoryear{{Marino} et~al.,}{{Marino}
  et~al.}{2015}]{Marino2015}
{Marino} A.~F.,  et~al., 2015, \mn@doi [\mnras] {10.1093/mnras/stv420}, \href
  {https://ui.adsabs.harvard.edu/abs/2015MNRAS.450..815M} {450, 815}

\bibitem[\protect\citeauthoryear{{Martin} et~al.,}{{Martin}
  et~al.}{2023}]{Martin2023}
{Martin} N.~F.,  et~al., 2023, \mn@doi [arXiv e-prints]
  {10.48550/arXiv.2308.01344}, \href
  {https://ui.adsabs.harvard.edu/abs/2023arXiv230801344M} {p. arXiv:2308.01344}

\bibitem[\protect\citeauthoryear{{Mastrobuono-Battisti}, {Khoperskov}, {Di
  Matteo}  \& {Haywood}}{{Mastrobuono-Battisti} et~al.}{2019}]{Mastrobuono2019}
{Mastrobuono-Battisti} A.,  {Khoperskov} S.,  {Di Matteo} P.,   {Haywood} M.,
  2019, \mn@doi [\aap] {10.1051/0004-6361/201834087}, \href
  {https://ui.adsabs.harvard.edu/abs/2019A&A...622A..86M} {622, A86}

\bibitem[\protect\citeauthoryear{{McKenzie} \& {Bekki}}{{McKenzie} \&
  {Bekki}}{2018}]{McKenzie2018}
{McKenzie} M.,  {Bekki} K.,  2018, \mn@doi [\mnras] {10.1093/mnras/sty1557},
  \href {https://ui.adsabs.harvard.edu/abs/2018MNRAS.479.3126M} {479, 3126}

\bibitem[\protect\citeauthoryear{{Monaco}, {Bellazzini}, {Ferraro}  \&
  {Pancino}}{{Monaco} et~al.}{2005}]{Monaco05}
{Monaco} L.,  {Bellazzini} M.,  {Ferraro} F.~R.,   {Pancino} E.,  2005, \mn@doi
  [\mnras] {10.1111/j.1365-2966.2004.08579.x}, \href
  {https://ui.adsabs.harvard.edu/abs/2005MNRAS.356.1396M} {356, 1396}

\bibitem[\protect\citeauthoryear{{Piatti}}{{Piatti}}{2023}]{Piatti2023}
{Piatti} A.~E.,  2023, \mn@doi [\mnras] {10.1093/mnrasl/slad098}, \href
  {https://ui.adsabs.harvard.edu/abs/2023MNRAS.tmpL..92P} {}

\bibitem[\protect\citeauthoryear{{Renzini} et~al.,}{{Renzini}
  et~al.}{2015}]{Renzini2015}
{Renzini} A.,  et~al., 2015, \mn@doi [\mnras] {10.1093/mnras/stv2268}, \href
  {https://ui.adsabs.harvard.edu/abs/2015MNRAS.454.4197R} {454, 4197}

\bibitem[\protect\citeauthoryear{{Soubiran}, {Brouillet}  \&
  {Casamiquela}}{{Soubiran} et~al.}{2022}]{Soubiran2022}
{Soubiran} C.,  {Brouillet} N.,   {Casamiquela} L.,  2022, \mn@doi [\aap]
  {10.1051/0004-6361/202142409}, \href
  {https://ui.adsabs.harvard.edu/abs/2022A&A...663A...4S} {663, A4}

\bibitem[\protect\citeauthoryear{{Taylor} et~al.,}{{Taylor}
  et~al.}{2022}]{Taylor2022}
{Taylor} D.~J.,  et~al., 2022, \mn@doi [\mnras] {10.1093/mnras/stac968}, \href
  {https://ui.adsabs.harvard.edu/abs/2022MNRAS.513.3429T} {513, 3429}

\bibitem[\protect\citeauthoryear{{Vasiliev} \& {Baumgardt}}{{Vasiliev} \&
  {Baumgardt}}{2021}]{Vasiliev2021}
{Vasiliev} E.,  {Baumgardt} H.,  2021, \mn@doi [\mnras]
  {10.1093/mnras/stab1475}, \href
  {https://ui.adsabs.harvard.edu/abs/2021MNRAS.505.5978V} {505, 5978}

\bibitem[\protect\citeauthoryear{{Yong} et~al.,}{{Yong}
  et~al.}{2014}]{Yong2014}
{Yong} D.,  et~al., 2014, \mn@doi [\mnras] {10.1093/mnras/stu806}, \href
  {https://ui.adsabs.harvard.edu/abs/2014MNRAS.441.3396Y} {441, 3396}

\makeatother
\end{thebibliography}



\bsp	
\label{lastpage}
\end{document}